\documentclass[12pt]{article}

\usepackage{amsfonts}
\usepackage[mathscr]{eucal}
\usepackage{bm}
\usepackage{cite}

\newcommand{\be}{\begin{equation}}
\newcommand{\ee}{\end{equation}}
\newcommand{\bea}{\begin{eqnarray}}
\newcommand{\eea}{\end{eqnarray}}

\newcommand{\p}[1]{(\ref{#1})}

\begin{document}

\begin{titlepage}

\vspace*{1.5cm}

\renewcommand{\thefootnote}{\dag}
\begin{center}

{\LARGE\bf $\mathcal{N}{=}\,4$ supersymmetric $\mathrm{U}(2)$-spin }

\vspace{0.45cm}

{\LARGE\bf hyperbolic Calogero-Sutherland model }

\vspace{1.5cm}

{\large\bf Sergey Fedoruk}
 \vspace{0.5cm}

{\it Bogoliubov Laboratory of Theoretical Physics, }\\
{\it Joint Institute for Nuclear Research,}\\
{\it 141980 Dubna, Moscow region, Russia} \\
\vspace{0.1cm}

{\tt fedoruk@theor.jinr.ru}

\vspace{1.5cm}

\end{center}

\vspace{1.2cm} \vskip 0.6truecm \nopagebreak

\begin{abstract}
\noindent
\qquad The $\mathcal{N}{=}\,4$ supersymmetric $\mathrm{U}(2)$-spin hyperbolic Calogero-Sutherland model
with odd matrix fields is examined.
Explicit form of the $\mathcal{N}{=}\,4$ supersymmetry generators is derived.
The Lax representation for the dynamics of the $\mathcal{N}{=}\,4$ hyperbolic $\mathrm{U}(2)$-spin Calogero-Sutherland system is found.
The reduction to the $\mathcal{N}{=}\,4$ supersymmetric spinless hyperbolic Calogero-Sutherland system is established.
\end{abstract}

\vspace{3cm}


\newpage

\end{titlepage}

\setcounter{footnote}{0}
\setcounter{equation}0
\section{Introduction}

One of the important developments in the study of the famous many-particle Calogero-Sutherland systems \cite{C,Su}  (see \cite{OP,Poly-rev} for reviews) is their generalization to supersymmetric cases.
Most of the researches in these directions have been devoted to supersymmetrization of the rational Calogero systems
(see, for example, the papers \cite{FrM,BHKVas,BGK,BGL,Wyl,GLP-2007,FIL08,KL-10,FI,FILS,KLS-18,KLS-18b,Feig,KLS-19,KL-20}
and the review  \cite{superc}).
Supersymmetric generalizations of the hyperbolic and trigonometric Calogero-Sutherland systems have been studied in a very limited
number of works
(see, for example, the papers \cite{SSuth,BrinkTurW,BorManSas,IoffeNee,DeLaMa,Serg,SergVes,Feig,KLS-19,KL-20} and references therein).

In a recent paper \cite{FIL19}, $\mathcal{N}{=}\, 2$ and $\mathcal{N}{=}\, 4$ supersymmetric generalizations
of the hyperbolic Calogero-Sutherland system were proposed using the gauging procedure \cite{DI-06-1,FIL08}
(see also the matrix description of the Calogero models in \cite{Poly-gauge,Gorsky,Poly-rev}).
In the paper \cite{Fed20} the $\mathcal{N}{=}\, 2$ hyperbolic Calogero-Sutherland model \cite{FIL19} was considered.
In this paper, the Hamiltonian analysis of the $\mathcal{N}{=}\, 4$ many-particle system obtained in \cite{FIL19} was studied in detail.

At the component level, the $\mathcal{N}{=}\, 4$ matrix model obtained in \cite{FIL19} is described by
the positive definite Hermitian $c$-number $(n{\times}n)$--matrix field
$X(t):=\|X_a{}^b(t)\|$, $({X_a{}^b})^* =X_b{}^a$, $\det X \neq 0$,
and the Hermitian $c$-number $(n{\times}n)$--matrix gauge field
$A(t):=\|A_a{}^b(t)\|$, $({A_a{}^b})^* =A_b{}^a$ ($a,b=1,\ldots ,n$).
In opposite to the  $\mathcal{N}{=}\,2$ case  \cite{Fed20}, the $\mathcal{N}{=}\, 4$ model
uses the complex odd $(n{\times}n)$--matrix field
$\Psi^i(t):=\|\Psi^i{}_a{}^b(t)\|$, $ \bar\Psi_i(t):=\|\bar\Psi_i{}_a{}^b(t)\|$,
$({\Psi^i{}_a{}^b})^* =\bar\Psi_i{}_b{}^a$,
and the complex $c$-number $\mathrm{U}(n)$-spinor field
$Z^k(t):=\|Z^k_a(t)\|$, $\bar Z_k(t):=\|\bar Z_k^a(t)\|$, $\bar Z_k^a = ({Z^k_a})^*$,
which have additional $\mathrm{SU}(2)$-spinor indices $i,k=1,2$.
This $\mathcal{N}{=}\, 4$ $n$-particle system is described by the on-shell component action
$
{\displaystyle S_{\rm matrix} = \int \mathrm{d}t \, L_{\rm matrix} }
$
with the Lagrangian (system II in  \cite{FIL19})
\begin{eqnarray}
\label{N2Cal-com}
L_{\rm matrix}  &  {=} & \frac12\ {\rm Tr}\Big( \,X^{-1}\nabla\! X \,X^{-1}\nabla\! X+ 2c\, A\Big)
\ + \ \frac{i}{2}\, \Big(\bar Z_k \nabla\! Z^k - \nabla\! \bar Z_k Z^k\Big)
\\ [5pt]
&&
+ \ \frac{i}{2}\ {\rm Tr} \Big( X^{-1}\bar\Psi_k X^{-1}\nabla \Psi^k - X^{-1}\nabla \bar\Psi_k X^{-1}\Psi^k \Big)
\nonumber\\ [5pt]
&&
+ \ \frac{1}{8}\, {\rm Tr} \Big( \{X^{-1}\Psi^i, X^{-1}\bar\Psi_i\}\, \{X^{-1}\Psi^{k} , X^{-1}\bar\Psi_{k}\} \Big)
\,,
\nonumber
\end{eqnarray}
where the quantity $c$ is a real constant and the
covariant derivatives are defined by
$\nabla\! X = \dot X +i\, [A, X]$ and $\nabla \Psi^k = \dot \Psi^k +i\, [A,\Psi^k]$,
$\nabla\! Z^k = \dot Z^k + iAZ^k$ and {\cal c.c.}

Despite the external similarity of the Lagrangian \p{N2Cal-com} with the $\mathcal{N}{=}\,2$ supersymmetric Lagrangian  \cite{Fed20}, the $\mathrm{SU}(2)$-spinor character of the Grassmann matrix quantities $\Psi^i$ and semi-dynamical even variables $Z^i$ leads to the distinctive properties of the $\mathcal{N}{=}\,4$ system under consideration.
First, using the $\mathrm{SU}(2)$-spinors  $Z^i$ leads to the $\mathcal{N}{=}\,4$ matrix system that is supersymmetric generalization of the $\mathrm{U}(2)$-spin hyperbolic Calogero-Sutherland system, and not the spinless hyperbolic Calogero-Sutherland system as in the $\mathcal{N}{=}\,2$ case \cite{Fed20}.
Second, due to the $\mathrm{SU}(2)$-spinor nature of the Grassmann matrix quantities $\Psi^i$, the $\mathcal{N}{=}\,4$ supercharges contain 
additional terms in odd variables, that were absent in the $\mathcal{N}{=}\,2$ case \cite{Fed20}.
This paper examines the $\mathcal{N}{=}\,4$ case in detail in order to identify these and other distinctive properties of this $\mathcal{N}{=}\,4$ system.

The plan of the paper is as follows.
In Section~2, the Hamiltonian formulation of the matrix system \p{N2Cal-com} is presented.
Partial gauge fixing eliminates purely gauge bosonic off-diagonal matrix fields and
yields a classically-equivalent system, whose bosonic limit is exactly
the multi-particle $\mathrm{U}(2)$-spin hyperbolic Calogero-Sutherland system.
Using the Noether procedure in Section~3 allows one to find the full set of $\mathcal{N}{=}\,4$ supersymmetry generators.
The Dirac brackets superalgebra of these generators is closed to first class constraints.
Section~4 is devoted to the construction of the Lax representation for the equation of motion
of the $\mathcal{N}{=}\,4$ supersymmetric $\mathrm{U}(2)$-spin hyperbolic Calogero-Sutherland system.
Section~5 presents the reduction of the considered $\mathrm{U}(2)$-spin system
that yields the $\mathcal{N}{=}\,4$ supersymmetric spinless hyperbolic Calogero-Sutherland system.
Section~6 contains summary and outlook.

\setcounter{equation}{0}
\section{Hamiltonian formulation}

Here we present the Hamiltonization of the matrix system (\ref{N2Cal-com}) with the $\mathrm{U}(n)$ gauge symmetry
and its partial gauge-fixing.

\subsection{Hamiltonian formulation of the matrix system}

The system with the Lagrangian (\ref{N2Cal-com}) is described by pairs of the phase variables
$(X_a{}^b, P_c{}^d)$, $(Z^i_a, \mathcal{P}_k^b)$, $(\bar Z_i^a, \bar\mathcal{P}^k_b)$,
$(\Psi^i{}_a{}^b, \Pi_k{}_c{}^d)$, $(\bar\Psi_i{}_a{}^b, \bar\Pi^k{}_c{}^d)$ with
the nonvanishing canonical Poisson brackets
\begin{equation}\label{PB-X}
\{X_a{}^b, P_c{}^d \}_{\scriptstyle{\mathrm{P}}} =  \delta_a^d \delta_c^b \,,\qquad
\{Z^i_a, \mathcal{P}_k^b \}_{\scriptstyle{\mathrm{P}}} =  \delta_a^b \delta^i_k \,,\quad
\{\bar Z_i^a, \bar\mathcal{P}^k_b \}_{\scriptstyle{\mathrm{P}}} =  \delta_b^a  \delta_i^k\,,
\end{equation}
\begin{equation}\label{PB-Psi}
\{\Psi^i{}_a{}^b, \Pi_k{}_c{}^d \}_{\scriptstyle{\mathrm{P}}} =  \delta_a^d \delta_c^b \delta^i_k \,, \quad
\{\bar\Psi_i{}_a{}^b, \bar\Pi^k{}_c{}^d \}_{\scriptstyle{\mathrm{P}}} =  \delta_a^d \delta_c^b  \delta_i^k \,.
\end{equation}
The phase variables  are subject to the primary constraints
\begin{equation}\label{const-Z}
G_k^a := \mathcal{P}_k^a - \frac{i}2 \, \bar Z_k^a \approx 0\,, \qquad
\bar G^k_a := \bar\mathcal{P}^k_a + \frac{i}2 \, Z^k_a \approx 0 \,,
\end{equation}
\begin{equation}\label{const-Psi}
\Upsilon_k{}_a{}^b := \Pi_k{}_a{}^b - \frac{i}2 \,(X^{-1}\bar\Psi_k X^{-1})_a{}^b\approx 0 \,, \qquad
\bar \Upsilon^k{}_a{}^b := \bar\Pi^k{}_a{}^b - \frac{i}2 \,(X^{-1} \Psi^k X^{-1})_a{}^b\approx 0 \,.
\end{equation}
Besides, the matrix momentum of $X_a{}^b$ has the form
\begin{equation}\label{P-X}
P_a{}^b = (X^{-1}\nabla X X^{-1})_a{}^b
\end{equation}
and the momenta of the coordinates $A_a{}^b$ are zero.

The canonical Hamiltonian of the system has the form
\begin{equation}\label{t-Ham}
H_{\rm matrix}=\ P_b{}^a \dot X_a{}^b + \mathcal{P}_k^a \dot Z^k_a + \bar\mathcal{P}^k_a \dot{\bar Z}_k^a +
\Pi_k{}_b{}^a \dot\Psi^k{}_a{}^b + \bar\Pi^k{}_b{}^a\dot{\bar\Psi}_k{}_a{}^b - L_{\rm matrix}\ =\ H+ {\rm Tr}\big(A F \big)\,,
\end{equation}
where the first term
has the following form
\begin{equation}
\label{Ham-matrix1}
H  \  = \
\frac12\,{\rm Tr}\Big(XPXP\Big)\ - \ \frac{1}{8}\, {\rm Tr} \Big( \{X^{-1}\Psi^i, X^{-1}\bar\Psi_i\}\, \{X^{-1}\Psi^{k} , X^{-1}\bar\Psi_{k}\} \Big)
\end{equation}
and the second term ${\rm Tr}\big(A F \big)$ uses the quantities
\begin{equation}\label{F-constr}
F_a{}^b := i[P,X]_a{}^b + Z^k_a\bar Z_k^b-\frac12\,\{X^{-1}\Psi^k, X^{-1}\bar\Psi_k \}_a{}^b
-\frac12\,\{\Psi^k X^{-1}, \bar\Psi_k X^{-1} \}_a{}^b - c\,\delta_a{}^b\,.
\end{equation}
Vanishing momenta of $A_a{}^b$ indicate that
quantities  \p{F-constr} are the secondary constraints
\begin{equation}\label{F-constr1}
F_a{}^b \approx 0\,.
\end{equation}
The variables $A_a{}^b$ in the Hamiltonian  \p{t-Ham} play the role of the Lagrange multipliers for these constraints.

The constraints \p{const-Z} and \p{const-Psi} possess the following nonzero Poisson brackets:
\begin{equation}\label{PB-const-2}
\{ G_i^a , \bar G^k_b \}_{\scriptstyle{\mathrm{P}}} =-i\delta^a_b \delta^k_i\,,\qquad
\{ \Upsilon_i{}_a{}^b , \bar \Upsilon^k{}_c{}^d \}_{\scriptstyle{\mathrm{P}}} =-iX^{-1}_{\ \ a}{}^d X^{-1}_{\ \ c}{}^b  \delta^k_i
\end{equation}
and are the second class constraints.
Using the Dirac brackets for the constraints \p{const-Z}, \p{const-Psi}
\begin{eqnarray}\label{DB-const-2}
\{ A,B\}_{\scriptstyle{\mathrm{D}}} &=&
\{ A,B\}_{\scriptstyle{\mathrm{P}}}
\ +\,i \{A  , G_k^a \}_{\scriptstyle{\mathrm{P}}} \{ \bar G^k_a , B \}_{\scriptstyle{\mathrm{P}}} -i \{ A , \bar G^k_a \}_{\scriptstyle{\mathrm{P}}}\{ G_k^a , B \}_{\scriptstyle{\mathrm{P}}} \\ [6pt]
&&
-i \{A  , \Upsilon_k{}_a{}^b \}_{\scriptstyle{\mathrm{P}}}X_b{}^c X_d{}^a\{ \bar \Upsilon^k{}_c{}^d, B \}_{\scriptstyle{\mathrm{P}}} -i
\{ A , \bar \Upsilon^k{}_a{}^b \}_{\scriptstyle{\mathrm{P}}}X_b{}^c X_d{}^a\{ \Upsilon_k{}_c{}^d , B \}_{\scriptstyle{\mathrm{P}}}
\,, \nonumber
\end{eqnarray}
we eliminate the momenta $\mathcal{P}_k^a$, $\bar\mathcal{P}^k_a$, $\Pi_k{}_a{}^b$, $\bar\Pi^k{}_a{}^b$.
The nonvanishing  Dirac brackets of residual phase variables take the form
\begin{equation}\label{DB-X}
\{X_a{}^b, P_c{}^d \}_{\scriptstyle{\mathrm{D}}} =  \delta_a^d \delta_c^b \,,
\end{equation}
\begin{equation}\label{DB-P}
\begin{array}{rcl}
\{P_a{}^b, P_c{}^d \}_{\scriptstyle{\mathrm{D}}} &= &  -\frac{i}4\, [X^{-1}(\Psi^k X^{-1}\bar\Psi_k + \bar\Psi_k X^{-1}\Psi^k)X^{-1}]_a{}^d X^{-1}_{\ \ c}{}^b \\ [5pt]
&& +\, \frac{i}4\, X^{-1}_{\ \ a}{}^d [X^{-1}(\Psi^k X^{-1}\bar\Psi_k + \bar\Psi_k X^{-1}\Psi^k)X^{-1}]_c{}^b \,,
\end{array}
\end{equation}
\begin{equation}\label{DB-Z}
\{Z^i_a, \bar Z_k^b \}_{\scriptstyle{\mathrm{D}}} =  -i\delta_a^b \delta_k^i\,,
\qquad
\{\Psi^i{}_a{}^b, \bar\Psi_k{}_c{}^d \}_{\scriptstyle{\mathrm{D}}} =  -iX_a{}^d X_c{}^b \delta_k^i \,,
\end{equation}
\begin{equation}\label{DB-PPs}
\begin{array}{rcl}
\{\Psi^k{}_a{}^b, P_c{}^d \}_{\scriptstyle{\mathrm{D}}} &=&  \frac{1}2\, \delta_a^d (X^{-1}\Psi^k )_c{}^b + \frac{1}2\, \delta_c^b (\Psi^k X^{-1})_a{}^d\,,\\ [5pt]
\{\bar\Psi_k{}_a{}^b, P_c{}^d \}_{\scriptstyle{\mathrm{D}}} &=&  \frac{1}2\, \delta_a^d (X^{-1}\bar\Psi_k )_c{}^b + \frac{1}2\, \delta_c^b (\bar\Psi_k X^{-1})_a{}^d\,.
\end{array}
\end{equation}

The residual constraints $F_a{}^b=(F_b{}^a)^*$, defined in  \p{F-constr1}, form the $u(n)$ algebra with respect to the Dirac brackets \p{DB-const-2}:
\begin{equation}\label{DB-FF}
\{F_a{}^b, F_c{}^d \}_{\scriptstyle{\mathrm{D}}} =-i \delta_a{}^d F_c{}^b + i \delta_c{}^b F_a{}^d \,.
\end{equation}
So the constraints \p{F-constr}, \p{F-constr1} are the first class ones and generate local $\mathrm{U}(n)$ transformations
\begin{equation}\label{trans-C}
\delta C= \sum_{a,b} \alpha_b{}^a \{C, F_a{}^b \}_{\scriptstyle{\mathrm{D}}}
\end{equation}
of an arbitrary phase variable $C$ where $\alpha_a{}^b(\tau)=(\alpha_b{}^a(\tau))^*$ are the local parameters.
These transformations of the primary phase variables have the form
\begin{equation}\label{Un-tran}
\begin{array}{c}
\delta X_a{}^b =-i[\alpha,X]_a{}^b \,, \quad \delta P_a{}^b =-i[\alpha,P]_a{}^b\,,\quad
\delta Z_a^k =-i(\alpha Z^k)_a \,, \quad
\delta  \bar Z_k{}_a =i(\bar Z_k \alpha)^a\,, \\ [7pt]
\delta \Psi^k{}_a{}^b = -i[\alpha,\Psi^k]_a{}^b \,, \quad
\delta \bar\Psi_k{}_a{}^b = -i[\alpha,\bar\Psi_k]_a{}^b  \,.
\end{array}
\end{equation}

\subsection{Hamiltonian formulation of partial gauge-fixing of the matrix system}

The gauges $X_a{}^b\,{=}\,0$ at $a\,{\neq}\,b$ fix the local transformations \p{Un-tran} with the parameters $\alpha_a{}^b(\tau)$, $a{\neq}b$
generated by the off-diagonal constraints $F_a{}^b \,{\approx}\,0$, $a\,{\neq}\,b$ in the set  \p{F-constr}, \p{F-constr1}.
This gauge fixing takes the form \cite{FIL08,FIL19,Fed20}
\begin{equation}\label{x-fix}
x_a{}^b\approx 0
\end{equation}
if we apply the expansions
\begin{equation}\label{XP-exp}
X_a{}^b =x_a \delta_a{}^b + x_a{}^b\,,
\qquad
P_a{}^b = \mathrm{p}_a \delta_a{}^b + \mathrm{p}_a{}^b\,,
\end{equation}
where $x_a{}^b$ and $\mathrm{p}_a{}^b$ represent the off-diagonal matrix quantities.
In addition, using
the constraints $F_a{}^b\,{\approx}\,0$, $a\,{\neq}\,b$, we express the momenta $\mathrm{p}_a{}^b$ through the remaining  phase variables:
\begin{equation}\label{p-exp}
\mathrm{p}_a{}^b= -\frac{i\,Z^k_a\bar Z_k^b}{x_a-x_b}+\frac{i\,(x_a+x_b)\,\{\Phi^k,\bar\Phi_k\}_a{}^b}{2(x_a-x_b)\sqrt{x_ax_b}}\,,
\end{equation}
where we use the odd matrix variables $\Phi^k{}_a{}^b$, $\bar\Phi_k{}_a{}^b=(\Phi^k{}_b{}^a)^*$ defined by
\begin{equation}\label{Phi-def}
\Phi^k{}_a{}^b:= \frac{\Psi^k{}_a{}^b}{\sqrt{x_ax_b}}\,,\qquad
\bar\Phi_k{}_a{}^b:= \frac{\bar\Psi_k{}_a{}^b}{\sqrt{x_ax_b}}\,.
\end{equation}
Thus, the partial gauge fixing conditions \p{x-fix} and \p{p-exp} remove the variables
$x_a{}^b$ and $\mathrm{p}_a{}^b$.

As a result, after the partial gauge fixing, phase space of the considered system is defined
by $2n$ even real variables $x_a$, $\mathrm{p}_a$, $2n$ even complex variables $Z^i_a$ and $2n^2$ odd complex variables $\Phi^i{}_a{}^b$.
Their nonvanishing Dirac brackets are
\begin{eqnarray}\label{DB-xp}
\{x_a, \mathrm{p}_b \}^{'}_{\scriptstyle{\mathrm{D}}} &=&  \delta_{ab} \,,
\\ [6pt]
\label{DB-Z1}
\{Z^i_a, \bar Z_k^b \}^{'}_{\scriptstyle{\mathrm{D}}} &=&  -i\,\delta_a^b \delta_k^i\,,
\\ [6pt]
\label{DB-Ph}
\{\Phi^i{}_a{}^b, \bar\Phi_k{}_c{}^d \}^{'}_{\scriptstyle{\mathrm{D}}} &=&  -i\,\delta_a^d \delta_c^b \delta_k^i \,.
\end{eqnarray}
In contrast to \p{DB-P} and \p{DB-PPs} the momenta $\mathrm{p}_a$ commute with each other and with the Grassmannian quantities $\Phi^k{}_a{}^b$.
Moreover, due to \p{DB-Ph}, the odd variables $\Phi^k{}_a{}^b$ and $\bar\Phi_i{}_b{}^a$ form canonical pairs
(compare with \p{DB-Z}).

In the Hamiltonian \p{Ham-matrix1} the momenta $\mathrm{p}_a$
are presented in the term $\sum_{a}(x_a \mathrm{p}_a)^2/2$.
Let us represent this term in standard form for particle kinetic energy.
For this we introduce the phase variables
\begin{equation}\label{p-q}
q_a=\log x_a \,,\quad p_{\,a}=x_a \mathrm{p}_a \,,\qquad \{q_a,p_{\,b}\}^{'}_{\scriptstyle{\mathrm{D}}} =\delta_{ab}\,.
\end{equation}
In these variables and \p{Phi-def} and after the gauge-fixing \p{x-fix}, \p{p-exp},
the Hamiltonian \p{Ham-matrix1} takes the form
\begin{equation}\label{Ham-fix}
\mathrm{H}  \ = \
\frac12\,\sum_{a}p_a p_a \ + \
\frac18\,\sum_{a\neq b} \frac{R_a{}^bR_b{}^a}{\sinh^2 \Big({\displaystyle\frac{q_a-q_b}{2}}\Big)}
\ - \, \frac{1}{8}\ {\rm Tr} \Big( \{\Phi^i, \bar\Phi_i\}\{ \Phi^k, \bar\Phi_k\} \Big)
\,,
\end{equation}
where
\begin{equation}\label{T-def}
R_a{}^b := Z^k_a\bar Z_k^b- \cosh\left(\frac{q_a-q_b}{2}\right)\{ \Phi^k, \bar\Phi_k \}_a{}^b\,.
\end{equation}

The residual first class constraints in the set \p{F-constr}, \p{F-constr1} are $n$ diagonal constraints
\begin{equation}\label{F-constr-d}
F_a := F_a{}^a =R_a{}^a -c= Z^k_a\bar Z_k^a- \{ \Phi^k, \bar\Phi_k \}_a{}^a - c\approx 0\qquad\mbox{(no summation over $a$)}\,,
\end{equation}
which form an abelian algebra with respect to the Dirac brackets \p{DB-Ph}
\begin{equation}
\label{DB-constr1}
\{F_a , F_b \}^{'}_{\scriptstyle{\mathrm{D}}} =  0
\end{equation}
and generate the $[\mathrm{U}(1)]^n$ gauge transformations of $Z^k_a$ and $\Phi^k{}_a{}^b$ with the local
parameters $\gamma_a(t)$:
\begin{equation}\label{b-Ab}
Z^k_a \rightarrow \, \mathrm{e}^{i\gamma_a} Z^k_a \,, \quad \bar Z_k^a  \rightarrow \,
\mathrm{e}^{-i\gamma_a}
\bar Z_k^a  \qquad (\mbox{no
sum over}\; a)\,,
\end{equation}
\begin{equation}\label{Psi-Ab}
\Phi^k{}_a{}^b \rightarrow \, \mathrm{e}^{i\gamma_a} \Phi^k{}_a{}^b \mathrm{e}^{-i\gamma_b}\,, \quad
\bar\Phi_k{}_a{}^b \rightarrow \, \mathrm{e}^{i\gamma_a} \bar\Phi_k{}_a{}^b  \mathrm{e}^{-i\gamma_b}\qquad (\mbox{no
sums over}\; a,b)\,.
\end{equation}

Similarly to \p{XP-exp}, we can use the expansions of the Grassmannian matrix quantities \p{Phi-def}
in the diagonal and off-diagonal parts:
\begin{equation}\label{Phi-exp}
\Phi^k{}_a{}^b =\varphi^k_a \delta_a{}^b + \phi^k{}_a{}^b\,,
\qquad
\bar\Phi_k{}_a{}^b =\bar\varphi_k{}_a \delta_a{}^b + \bar\phi_k{}_a{}^b\,,
\end{equation}
where $\phi^k{}_a{}^a=\bar\phi_k{}_a{}^a=0$ at the fixed index $a$.
The Dirac brackets \p{DB-Ph} of the diagonal quantities $\varphi^k_a$, $\bar\varphi_k{}_a$ and
the off-diagonal ones $\phi^k{}_a{}^b$, $\bar\phi_k{}_a{}^b$ have the form
\begin{equation}
\label{DB-Ph1}
\{\varphi^i_a , \bar\varphi_k{}_b \}^{'}_{\scriptstyle{\mathrm{D}}} =  -i\,\delta_{a b}\delta^i_k\,,\qquad
\{\phi^i{}_a{}^b, \bar\phi_k{}_c{}^d \}^{'}_{\scriptstyle{\mathrm{D}}} =  -i\,\delta_a^d \delta_c^b\delta^i_k \,.
\end{equation}
The constraints \p{F-constr-d} involve only the off-diagonal fermions $\phi$, $\bar\phi$:
\begin{equation}\label{F-constr-d1}
F_a = Z^k_a\bar Z_k^a- \{ \phi^k, \bar\phi_k \}_a{}^a - c\approx 0\qquad\mbox{(no summation over $a$)}\,.
\end{equation}
In the variables $\varphi$, $\bar\varphi$, $\phi$, $\bar\phi$ the Hamiltonian \p{Ham-fix} takes the form
\begin{eqnarray}
\mathrm{H} &=& \frac12\,\sum_{a}p_a p_a \ + \
\frac18\,\sum_{a\neq b} \frac{\bar Z_i^a Z^k_a \ \bar Z_k^b Z^i_b}{\sinh^2 \Big({\displaystyle\frac{q_a-q_b}{2}}\Big)}
\nonumber\\
&&
+\,
\frac14\,\sum_{a\neq b} \frac{\coth \Big({\displaystyle\frac{q_a-q_b}{2}}\Big) }
{\sinh \Big({\displaystyle\frac{q_a-q_b}{2}}\Big)}\,Z^i_a \bar Z_i^b
\Big[(\varphi^k_a-\varphi^k_b)\bar\phi_k{}_b{}^a+
(\bar\varphi_k{}_a-\bar\varphi_k{}_b)\phi^k{}_b{}^a - \{\phi^k,\bar\phi_k \}_b{}^a\Big]
\nonumber\\
&&
+\,
\frac18\,\sum_{a\neq b}
\frac{1}
{\sinh^2 \Big({\displaystyle\frac{q_a-q_b}{2}}\Big)}\,
\Big[(\varphi^i_a-\varphi^i_b)(\varphi^k_a-\varphi^k_b)\bar\phi_i{}_a{}^b\bar\phi_k{}_b{}^a
+ (\bar\varphi_i{}_a-\bar\varphi_i{}_b)(\bar\varphi_k{}_a-\bar\varphi_k{}_b)\phi^i_a{}^b\phi^k_b{}^a
\nonumber \\
&&\qquad\qquad\qquad \qquad\qquad\quad
+\ 2(\varphi^i_a-\varphi^i_b)(\bar\varphi_k{}_a-\bar\varphi_k{}_b)\bar\phi_i{}_a{}^b \phi^k{}_b{}^a
+ \{\phi^i,\bar\phi_i \}_a{}^b \{\phi^k,\bar\phi_k \}_b{}^a
\nonumber \\
&&\qquad\qquad\qquad \qquad\qquad\quad
+\ 2(\bar\varphi_i{}_a-\bar\varphi_i{}_b)\phi^i{}_a{}^b \{\phi^k,\bar\phi_k \}_b{}^a
+2(\varphi^i_a-\varphi^i_b)\bar\phi_i{}_a{}^b \{\phi^k,\bar\phi_k \}_b{}^a \Big]
\nonumber\\
&&
-\,
\frac18\,\sum_{a} \{\phi^i,\bar\phi_i \}_a{}^a \{\phi^k,\bar\phi_k \}_a{}^a\,.
\label{Ham-fix1}
\end{eqnarray}

In the bosonic limit the Hamiltonian \p{Ham-fix1} takes the form
\begin{equation}\label{H-bose-lim}
\mathrm{H}_{bose} \ =\ \frac12\,\sum_{a}p_a p_a  +
\frac18\,\sum_{a\neq b} \frac{S_a{}_i{}^k  S_b{}_k{}^i}{\sinh^2 \Big({\displaystyle\frac{q_a-q_b}{2}}\Big)}
\,,
\end{equation}
where the quantities
\begin{equation}
\label{S-Z-def}
S_a{}_i{}^k:= {\bar Z}{}_i^{a}Z^k_{a}
\end{equation}
at all values $a$ form the $u(2)$ algebras with respect to the Dirac brackets:
\begin{equation}
\label{S-S-Dir}
\{S_a{}_i{}^k , S_b{}_j{}^l \}^{'}_{\scriptstyle{\mathrm{D}}} =  -i\,\delta_{a b}\left(\delta^k_j S_a{}_i{}^l -\delta^l_i S_a{}_j{}^k\right)\,.
\end{equation}
Thus, the Hamiltonian \p{H-bose-lim} has the form
\begin{equation}\label{H-bose-lim1}
\mathrm{H}_{bose} \ =\ \frac12\,\sum_{a}p_a p_a  +
\frac18\,\sum_{a\neq b} \frac{\mathrm{Tr}(S_a S_b)}{\sinh^2 \Big({\displaystyle\frac{q_a-q_b}{2}}\Big)}
\end{equation}
and is same as the Hamiltonian of the $\mathrm{U}(2)$-spin hyperbolic Calogero-Sutherland $A_{n-1}$-root system \cite{GH-84,W-85,Poly-rev}.

Derivation of this many-particle spin system in the ${\mathcal N}{=}\,4$ case is the result of using semi-dynamical
$\mathrm{SU}(2)$-spinor variables, which are the field components of the {\bf (4,4,0)} multiplets \cite{FIL19}.
In contrast to the ${\mathcal N}{=}\,4$ case considered here, the use of semi-dynamical scalar variables in the ${\mathcal N}{=}\,2$ case produces
``a less rich'' supersymmetric system, namely the ${\mathcal N}{=}\,2$ spinless hyperbolic Calogero-Sutherland system \cite{Fed20}.

\setcounter{equation}{0}
\section{${\mathcal N}{=}\,4$ supersymmetry  generators}

As discussed in Sect.\,1, the system (\ref{N2Cal-com}) considered here was derived from the ${\mathcal N}{=}\,4$ superfield model \cite{FIL19}.
Therefore, it is invariant under ${\mathcal N}{=}\,4$ supersymmetry transformations of the matrix component fields:
\begin{equation}\label{tr-susy}
\begin{array}{rcl}
\delta X &=& -\varepsilon_i \Psi^i +  \bar\varepsilon^{\,i} \bar\Psi_i\,, \\ [6pt]
\delta \Psi^{i} &=& i\,\bar\varepsilon^{\,i}\,\nabla X + \bar\varepsilon_{k} X\Big[ X^{-1}\Psi^{(i} ,X^{-1}\bar\Psi^{k)} \Big]\,,\\ [6pt]
\delta \bar\Psi_{i} &=&  i\, \varepsilon_{i}\,\nabla X +  \varepsilon^{\,k} X\Big[ X^{-1}\Psi_{(i} ,X^{-1}\bar\Psi_{k)} \Big]\,, \\ [6pt]
\delta Z^{i}&=&0\,,\quad \delta \bar Z_{i}\ =\ 0\,,\qquad \delta A\ =\ 0\,,
\end{array}
\end{equation}
where $\varepsilon_k$, $\bar\varepsilon^k=(\varepsilon_k)^*$ is two complex Grassmannian parameters.
These transformations are generated by the following Noether charges:
\begin{equation}\label{Q-matrix}
\begin{array}{rcl}
Q^i &=& {\displaystyle  {\rm Tr}\,\Big( P\Psi^i \ + \ \frac{i}{2}\, X^{-1}\bar\Psi^i X^{-1}\Psi_kX^{-1}\Psi^k \Big), } \\ [7pt]
\bar Q_i &=& {\displaystyle  {\rm Tr}\,\Big( P\bar\Psi_i \ + \ \frac{i}{2}\, X^{-1}\Psi_i X^{-1}\bar\Psi^kX^{-1}\bar\Psi_k  \Big) },
\end{array}
\end{equation}
where the matrix momentum $P_a{}^b$ is presented in \p{P-X}.
The supercharges \p{Q-matrix} and the Hamiltonian $H$ defined in  \p{Ham-matrix1}
form the ${\mathcal N}{=}\,4$ $d{=}\,1$ superalgebra
\begin{equation}\label{N2-susy-matrix}
\{ Q^i, \bar Q_j \}_{\scriptstyle{\mathrm{D}}} = -2i\,H\,\delta^i_j\,,\qquad
\{ Q^i, H \}_{\scriptstyle{\mathrm{D}}}=\{ \bar Q_i, H \}_{\scriptstyle{\mathrm{D}}}=0
\end{equation}
 with respect to the Dirac brackets \p{DB-X}-\p{DB-PPs}.

Putting the partial gauge fixing conditions \p{x-fix}, \p{p-exp} in expressions \p{Q-matrix}
and going to the variables \p{Phi-def}, \p{p-q}, we obtain the ${\mathcal N}{=}\,4$ supersymmetry generators
\begin{equation}\label{Q}
\begin{array}{rcl}
\mathrm{Q}^{\,i} &=& {\displaystyle  \sum\limits_{a} p_a \Phi^i{}_a{}^a \ -\ \frac{i}{2}\sum\limits_{a\neq b}
\frac{ R_a{}^b \Phi^i{}_b{}^a}{\sinh\Big({\displaystyle\frac{q_a-q_b}{2}}\Big)}
\ +\ \frac{i}{2}\sum\limits_{a, b}\, [\Phi^k, \bar\Phi_k]_a{}^b\Phi^i{}_b{}^a \,, } \\ [8pt]
\bar{\mathrm{Q}}_{\,i} &=& {\displaystyle  \sum\limits_{a} p_a \bar\Phi_i{}_a{}^a\ -\ \frac{i}{2}\sum\limits_{a\neq b}
\frac{ R_a{}^b \bar\Phi_i{}_b{}^a}{\sinh\Big({\displaystyle\frac{q_a-q_b}{2}}\Big)}
\ -\ \frac{i}{2}\sum\limits_{a, b}\, [\Phi^k, \bar\Phi_k]_a{}^b\bar\Phi_i{}_b{}^a }
\end{array}
\end{equation}
for the partial gauge fixing system,
which is described by the Hamiltonian \p{Ham-fix} and the first class constraints \p{F-constr-d}.
Using the Grassmannian variables $\varphi^i_a$, $\bar\varphi_i{}_a$,
$\phi^i{}_a{}^b$, $\bar\phi_i{}_a{}^b$, defined in \p{Phi-exp}, we cast the generators \p{Q} in the form
\begin{eqnarray}
\label{Q1}
\mathrm{Q}^{\,i} &= &\sum\limits_{a} p_a \varphi^i_a-\frac{i}{2}\sum\limits_{a\neq b}
\frac{ Z^k_a\bar Z_k^b\, \phi^i{}_b{}^a}{\sinh\Big({\displaystyle\frac{q_a-q_b}{2}}\Big)} \\
&&  +\,\frac{i}{2}\sum\limits_{a\neq b}
\coth \Big({\displaystyle\frac{q_a-q_b}{2}}\Big)
\Big[ (\bar\varphi_k{}_a-\bar\varphi_k{}_b)\phi^k{}_a{}^b  +(\varphi^k_a-\varphi^k_b)\bar\phi_k{}_a{}^b +\{\phi^k,\bar\phi_k\}_a{}^b
\Big]\,\phi^i{}_b{}^a
\nonumber\\
&&
+\,\frac{i}{2}\,\Big[
\sum\limits_{a\neq b} \Big((\varphi_k{}_a+\varphi_k{}_b)\phi^k{}_b{}^a \bar\phi^i{}_a{}^b +
\phi_k{}_a{}^b \phi^k{}_b{}^a \bar\varphi^i_a \Big)
+\!\!\!\!
\sum\limits_{a\neq b\neq c\neq a} \!\!\!\! \phi_k{}_a{}^b \phi^k{}_b{}^c \bar\phi^i{}_c{}^a   +
\sum\limits_{a} \varphi_k{}_a\varphi^k_a\bar\varphi^i_a \Big]\,,
\nonumber
\\
\label{bQ1}
\bar{\mathrm{Q}}_{\,i} &= &\sum\limits_{a} p_a \bar\varphi_i{}_a-\frac{i}{2}\sum\limits_{a\neq b}
\frac{ Z^k_a\bar Z_k^b\, \bar\phi_i{}_b{}^a}{\sinh\Big({\displaystyle\frac{q_a-q_b}{2}}\Big)} \\
&& +\,\frac{i}{2}\sum\limits_{a\neq b}
\coth\Big({\displaystyle\frac{q_a-q_b}{2}}\Big)
\Big[ (\bar\varphi_k{}_a-\bar\varphi_k{}_b)\phi^k{}_a{}^b  +(\varphi^k_a-\varphi^k_b)\bar\phi_k{}_a{}^b +\{\phi^k,\bar\phi_k\}_a{}^b
\Big]\,\bar\phi_i{}_b{}^a
\nonumber\\
&&
+\,\frac{i}{2}\,\Big[
\sum\limits_{a\neq b} \Big(\phi_i{}_a{}^b \bar\phi^k{}_b{}^a (\bar\varphi_k{}_a+\bar\varphi_k{}_b) +
\varphi_i{}_a \bar\phi^k{}_a{}^b \bar\phi_k{}_b{}^a \Big)
+\!\!\!\!
\sum\limits_{a\neq b\neq c\neq a} \!\!\!\! \phi_i{}_a{}^b \bar\phi^k{}_b{}^c \bar\phi_k{}_c{}^a   +
\sum\limits_{a} \varphi_i{}_a \bar\varphi^k_a\bar\varphi_k{}_a\Big]\,.
\nonumber
\end{eqnarray}
Taking into account the Dirac brackets \p{DB-Ph}, \p{p-q} and
\begin{equation}\label{R-alg}
\begin{array}{rcl}
\{R_a{}^b, R_c{}^d \}^{'}_{\scriptstyle{\mathrm{D}}}&=& -i\Big(\delta_a^d R_c{}^b-\delta_c^b R_a{}^d \Big)\\
&& -i\,
\sinh\Big({\displaystyle\frac{q_a-q_b}{2}}\Big)\sinh\Big({\displaystyle\frac{q_c-q_d}{2}}\Big)
\Big(\delta_a^d\{ \Phi^k, \bar\Phi_k \}_c{}^b-\delta_c^b\{ \Phi^k, \bar\Phi_k \}_a{}^d\Big)\,,
\end{array}
\end{equation}
we find that the supercharges $\mathrm{Q}^i$, $\bar \mathrm{Q}_i$ defined in  \p{Q}
form the ${\mathcal N}{=}\,4$ superalgebra
\begin{eqnarray}\label{DB-QQ}
\{\mathrm{Q}^i, \mathrm{Q}^k \}^{'}_{\scriptstyle{\mathrm{D}}} &=&  -\frac{i}{4}\sum\limits_{a\neq b}
\frac{ \phi^{(i}{}_a{}^b \phi^{k)}{}_b{}^a}{\sinh^2\Big({\displaystyle\frac{q_a-q_b}{2}}\Big)}\,\Big( F_a-F_b\Big) \,,
\\ [6pt]
\label{DB-QbQ}
\{ \mathrm{Q}^i , \bar \mathrm{Q}_k \}^{'}_{\scriptstyle{\mathrm{D}}} &=&  -2i\,\mathrm{H}\,\delta^i_k  -\frac{i}{4}\sum\limits_{a\neq b}
\frac{ \phi^{i}{}_a{}^b \bar\phi_{k}{}_b{}^a}{\sinh^2\Big({\displaystyle\frac{q_a-q_b}{2}}\Big)}\,\Big( F_a-F_b\Big) \,,
\\ [6pt]
\label{DB-HQ}
\{ \mathrm{Q}^i, \mathrm{H} \}^{'}_{\scriptstyle{\mathrm{D}}} &=&
-\frac{1}{8}\sum\limits_{a\neq b}
\frac{ R_a{}^b \phi^{i}{}_b{}^a}{\sinh^3\Big({\displaystyle\frac{q_a-q_b}{2}}\Big)}\,\Big( F_a-F_b\Big)
\end{eqnarray}
and {\cal c.c.}, where the Hamiltonian $\mathrm{H}$ and the constraints $F_a\approx0$ are given in \p{Ham-fix} and \p{F-constr-d}.
Thus, the quantities  $\mathrm{H}$, $\mathrm{Q}^i$, $\bar \mathrm{Q}_i$,
defined in \p{Ham-fix}, \p{Q},
form the $\mathcal{N}{=}\,4$ superalgebra with respect to the Dirac brackets
on the shell of the first class constraints \p{F-constr-d}.
Moreover, the generators $\mathrm{H}$, $\mathrm{Q}^i$, $\bar \mathrm{Q}_i$
are gauge invariant: they have the vanishing Dirac brackets with the first class constraints  \p{F-constr-d},
\begin{equation}
\label{DB-constr1-Q}
\{\mathrm{Q}^i , F_a \}^{'}_{\scriptstyle{\mathrm{D}}} =  \{\bar\mathrm{Q}_i , F_a \}^{'}_{\scriptstyle{\mathrm{D}}} =
\{ \mathrm{H} , F_a \}^{'}_{\scriptstyle{\mathrm{D}}} =0 \,.
\end{equation}

The form of the first two terms in expressions \p{Q} is similar to the ${\mathcal N}{=}\,2$ supercharges
presented in \cite{Fed20}.
But the last terms in the ${\mathcal N}{=}\,4$ supercharges \p{Q} were absent in the ${\mathcal N}{=}\,2$ case.
Their appearance is the result of the $\mathrm{SU}(2)$ spinor nature of Grassmann variables in the ${\mathcal N}{=}\,4$ case.
Moreover, the first and last terms  in the supercharges \p{Q1}, \p{bQ1}
\begin{equation} \label{Q-A-2n}
{\mathbb{Q}}^i = \sum_a \left( p_a\varphi^i_a  +\frac{i}{2}\, \varphi_k{}_a \varphi^k_a\bar\varphi^i_a\right) ,\qquad
\bar{\mathbb{Q}}_i =  \sum_a \left( p_a\bar\varphi_i{}_a  +\frac{i}{2}\, \varphi_i{}_a \bar\varphi^k_a\bar\varphi_k{}_a\right)
\end{equation}
contain only diagonal fermions $\varphi^i_a$, $\bar\varphi_i{}_a$ and possess the following Dirac brackets:
\begin{equation} \label{alg1-cl-An}
\{ {\mathbb{Q}}^i, \bar {\mathbb{Q}}_k \}_{\scriptstyle{\mathrm{D}}} = -2i\delta^i_k {\mathbb{H}}\,,\qquad
\{ {\mathbb{Q}}^i, {\mathbb{H}} \}_{\scriptstyle{\mathrm{D}}} = \{ \bar {\mathbb{Q}}_i, {\mathbb{H}} \}_{\scriptstyle{\mathrm{D}}} = 0\,,
\end{equation}
where
${\mathbb{H}}  = \frac12\sum_a  p_a^{\ 2}$.
Although supercharges \p{Q-A-2n} contain terms trilinear in fermions in contrast to the ${\mathcal N}{=}\,2$ case \cite{Fed20},
these quantities and ${\mathbb{H}}$ generate the ${\mathcal N}{=}\,4$ supersymmetric system
describing $n$ non-interacting free particles. This system is described by the ${\mathcal N}{=}\,4$
superfield Lagrangian $\mathscr{L}\sim \sum_a\log{\mathscr{X}_a}$ (see \cite{superc,IKL89,FIL12}).

It should also be noted that the terms of the supercharges \p{Q1}, \p{bQ1},
without the first and last terms \p{Q-A-2n}, describe the interaction of particles
and are zero when the off-diagonal matrix fermions $\phi^i{}_a{}^b$, $\bar\phi_i{}_a{}^b$ vanish.

Similarly to the ${\mathcal N}{=}\,2$ case \cite{Fed20}, we can make gauge-fixing for the residual $n$ real
first class constraints \p{F-constr-d} (or \p{F-constr-d1}).
However, in the considered ${\mathcal N}{=}\,4$ case, we have $2n$ complex spinor variables $Z^i_a$
in opposite to the ${\mathcal N}{=}\,2$ case with $n$ complex spinorial degrees of freedom  in the last case.
Thus, in the ${\mathcal N}{=}\,4$ case considered here the ${\mathcal N}{=}\,4$ multiparticle model possesses
$n$ complex semi-dynamical degrees of freedom in phase space and describes the ${\mathcal N}{=}\,4$ supersymmetrization
of the many-particle system which differs from the one in the ${\mathcal N}{=}\,2$ case. 
In Section\,5, we use the reduction that eliminates these semi-dynamical degrees of freedom in the ${\mathcal N}{=}\,4$ invariant way.

\setcounter{equation}{0}
\section{Lax representation}

Classical dynamics of the system with partial gauge-fixing considered here is defined by the total Hamiltonian
\begin{equation}\label{Ham-fix-t}
\mathrm{H}_{\mathrm{T}} \ = \ \mathrm{H}\ +\ \sum_a\lambda_aF_a \,,
\end{equation}
where the Hamiltonian $\mathrm{H}$ is defined in \p{Ham-fix} and $\lambda_a(t)$ are the Lagrange multipliers
for the first class constraints $F_a$, presented in \p{F-constr-d}.
A time derivative of an arbitrary phase variable $B(t)$ takes the form
\begin{equation}\label{der-B}
\dot B \ = \ \{ B, \mathrm{H}_{\mathrm{T}} \}^{'}_{\scriptstyle{\mathrm{D}}}  \,.
\end{equation}
Let us represent this dynamics in the Lax representation \cite{Lax}.

To do this, we introduce the $n{\times}n$ matrix
\begin{equation}
\label{L-matr}
L_{a}{}^{b} \ = \ p_a\, \delta_{a}{}^{b} \ - \ i\left( 1- \delta_{a}^{b} \right)
\frac{ R_a{}^b }{2\sinh\Big({\displaystyle\frac{q_a-q_b}{2}}\Big)}\,,
\end{equation}
whose evolution
\begin{equation}
\label{L-der-eq}
\dot L_{a}{}^{b} \ = \ \{ L_{a}{}^{b}, \mathrm{H}_{\mathrm{T}} \}^{'}_{\scriptstyle{\mathrm{D}}}
\end{equation}
is represented by the matrix commutator
\begin{equation}
\label{L-eq}
\dot L_{a}{}^{b} \ = \ -i [M+\Lambda,L]_{a}{}^{b}
-i\left( 1- \delta_{a}^{b} \right)\frac{ L_a{}^b\left(F_a-F_b\right)}{4\sinh^2\Big({\displaystyle\frac{q_a-q_b}{2}}\Big)} \,,
\end{equation}
where the matrices $M$ and  $\Lambda$ have the following form:
\begin{equation}
\label{M-matr}
M_{a}{}^{b} \ = \
\frac{1}{4}\,\{\Phi^k,\bar\Phi_k\}_{a}{}^{a}\delta_{a}{}^{b} \ + \ \frac{1}{4}\left( 1- \delta_{a}^{b} \right)
\left(\frac{ \cosh\Big({\displaystyle\frac{q_a-q_b}{2}}\Big)}
{\sinh^2\Big({\displaystyle\frac{q_a-q_b}{2}}\Big)}\,R_a{}^b +\{\Phi^k,\bar\Phi_k\}_{a}{}^{b}\right),
\end{equation}
\begin{equation}
\label{La-matr}
\Lambda_{a}{}^{b} \ = \ \lambda_{a}\, \delta_{a}{}^{b}
\end{equation}
and  $F_a$ are the constraints defined in  \p{F-constr-d1}.
The equations of motion
of the fermionic matrix variables $\Phi^i_{a}{}^{b}$, $\bar\Phi_i{}_{a}{}^{b}$
are also represented as commutators
\begin{equation}
\label{Ps-eq}
\begin{array}{rcccl}
\dot \Phi^i_{a}{}^{b} &=& \{ \Phi^i_{a}{}^{b}, \mathrm{H}_{\mathrm{T}} \}^{'}_{\scriptstyle{\mathrm{D}}} &=& -i [M+\Lambda,\Phi^i]_{a}{}^{b}
\,, \\ [7pt]
\dot {\bar\Phi}_i{}_{a}{}^{b} &=& \{ {\bar\Phi}_i{}_{a}{}^{b}, \mathrm{H}_{\mathrm{T}} \}^{'}_{\scriptstyle{\mathrm{D}}} &=& -i [M+\Lambda,{\bar\Phi_i}]_{a}{}^{b}
\end{array}
\end{equation}
with the same matrices $M$ and  $\Lambda$.

On the shell of the first class constraints \p{F-constr-d1} $F_a\approx 0$, equations \p{L-eq}, \p{Ps-eq} are actually the Lax equations
and yield the conserved charges  in a simple way.
So due to equations \p{L-eq}, \p{Ps-eq}, the trace
\begin{equation}
\label{J-f-def}
\mathcal{J} := {\mathrm{Tr}} (\mathcal{F})
\end{equation}
of any polynomial function
$\mathcal{F}(L,\Phi,\bar\Phi)$ of the matrix variables $L_{a}{}^{b}$, ${\Phi}^i_{a}{}^{b}$, ${\bar\Phi}_i{}_{a}{}^{b}$
is a conserved quantity on the shell of constraints \p{F-constr-d1}:
\begin{equation}
\label{F-conser}
\dot \mathcal{J} \approx 0\,.
\end{equation}
In particular, on the shell of constraints \p{F-constr-d1}, the traces
\begin{equation}
\label{Ik-def}
I_\mathrm{k} := {\mathrm{Tr}} (L^\mathrm{k})\,,\qquad \mathcal{I}^i_{\,\mathrm{k}} := {\mathrm{Tr}} (\Phi^i L^\mathrm{k-1})\,, \qquad \bar{\mathcal{I}}_{\,\mathrm{k}}{}_i := {\mathrm{Tr}} (\bar\Phi L^\mathrm{k-1})\,,\qquad \mathrm{k}=1,\ldots,n
\end{equation}
are conserved:
\begin{equation}
\label{Ik-eq}
\dot I_\mathrm{k} = \frac{i\mathrm{k}}{4}\sum\limits_{a\neq b}
\frac{ (L^{\mathrm{k}})_a{}^b }{\sinh^2\Big({\displaystyle\frac{q_a-q_b}{2}}\Big)}\,\Big( F_a-F_b\Big) \approx 0\,,
\qquad \dot {\mathcal{I}}^i_{\,\mathrm{k}} = 0\,,\qquad \dot {\bar{\mathcal{I}}}_{\,\mathrm{k}}{}_i = 0\,.
\end{equation}

The Hamiltonian \p{Ham-fix} and the supercharges \p{Q} have the form
\begin{equation}\label{Ham-fix-c}
\mathrm{H} = \frac12\,I_2 +J\,,\qquad \mathrm{Q}^i =\mathcal{I}^i_{\,2}+\mathcal{J}^i\,,
\qquad \bar \mathrm{Q}_i =\bar{\mathcal{I}}_{\,2}{}_i+\bar\mathcal{J}_i \,,
\end{equation}
where
\begin{equation}\label{Ham-fix3}
J := -\frac18 \,{\rm Tr}\Big(\{\Phi^i,\bar\Phi_i\}\{\Phi^k,\bar\Phi_k\}\Big)\,,
\quad \mathcal{J}^i := \frac{i}{2} \,{\rm Tr}\Big([\Phi^k,\bar\Phi_k]\Phi^i \Big)\,,\quad
\bar\mathcal{J}_i := -\frac{i}{2} \,{\rm Tr}\Big(\bar\Phi_i[\Phi^k,\bar\Phi_k] \Big)\,.
\end{equation}

The equations of motion
of the commuting spinning variables $Z^i_{a}$, $\bar Z_i^{a}$
are represented as
\begin{equation}
\label{Z-eq}
\begin{array}{rcccl}
\dot Z^i_{a} & = & \{ \Phi^i_{a}{}^{b}, \mathrm{H} \}^{'}_{\scriptstyle{\mathrm{D}}} & = &
-i \sum\limits_{ b} \left(A_{a}{}^{b} +\Lambda_{a}{}^{b}\right) Z^i_{b}\,,\\ [7pt]
\dot {\bar Z}{}_i^{a} & = & \{ {\bar\Phi}_i{}_{a}{}^{b}, \mathrm{H} \}^{'}_{\scriptstyle{\mathrm{D}}} & = & i \sum\limits_{ b} {\bar Z}{}_i^{b} \left(A_{b}{}^{a} +\Lambda_{b}{}^{a}\right),
\end{array}
\end{equation}
where the matrix $A$ has the form
\begin{equation}
\label{A-matr}
A_{a}{}^{b} \ = \    \left( 1- \delta_{a}^{b} \right)
\frac{ R_a{}^b} {4\sinh^2\Big({\displaystyle\frac{q_a-q_b}{2}}\Big)}
\end{equation}
and the matrix $\Lambda$ is defined in \p{La-matr}.
Due to \p{Z-eq} we obtain (see \p{S-Z-def})
\begin{equation}
\label{Z-inv}
\dot S_k{}^i = 0 \,,\qquad \mbox{where} \qquad S_k{}^i:=\sum\limits_{a}{\bar Z}{}_k^{a}Z^i_{a}   \,.
\end{equation}

It should be noted that the structure of the conserved charges in the considered supersymmetric system \p{F-conser}
is similar to the form of the charges in the trigonometric (non-matrix) supersymmetric system studied in \cite{DeLaMa}.

Deriving the Lax pair and finding the set of conserved charges \p{J-f-def} paves the way for analyzing the integrability
of the $\mathcal{N}{=}\,4$ supersymmetric system considered here.
Analysis of the superalgebra of conserved charges and integrability of
the considered many-particle supersymmetric system will be the subject of the next article.

\setcounter{equation}{0}
\section{Spinless hyperbolic Calogero-Sutherland system as a result of the reduction procedure}

Semi-dynamical variables have the following Dirac brackets with the total Hamiltonian \p{Ham-fix-t}, \p{Ham-fix}
\begin{equation}
\label{sd-H}
\{ H_T, Z_a^j \}^{'}_{\scriptstyle{\mathrm{D}}} =
{\displaystyle\frac{i}4\sum_{b(\neq a)} \frac{R_a{}^b Z_b^j}{\sinh^2\Big({\displaystyle\frac{q_a-q_b}{2}}\Big)}
+i\lambda_a Z_a^j}
\end{equation}
and with the supercharges \p{Q}
\begin{equation}
\label{sd-Q}
\{ Q^i, Z_a^j \}^{'}_{\scriptstyle{\mathrm{D}}} =
-{\displaystyle\frac12\sum_{b(\neq a)} \frac{\Phi^i{}_a{}^b Z_b^j}{\sinh\Big({\displaystyle\frac{q_a-q_b}{2}}\Big)}}\,, \qquad
\{ \bar Q_i, Z_a^j \}^{'}_{\scriptstyle{\mathrm{D}}} =
-{\displaystyle\frac12\sum_{b(\neq a)} \frac{  \bar\Phi_i{}_a{}^b Z_b^j}{\sinh\Big({\displaystyle\frac{q_a-q_b}{2}}\Big)}}\,.
\end{equation}
Therefore, the conditions
\begin{equation}
\label{red-Z}
Z_a^{j=2}=0\,, \qquad
\bar Z^a_{j=2} =0\,,\qquad \mbox{at all $a$}
\end{equation}
are invariant under the $\mathcal{N}{=}\,4$ supersymmetry transformations and we can use them as reduction conditions.
Similarly to \cite{KL-09}, the reduction \p{red-Z} implies the conditions
\begin{equation}
\label{red-gen}
S^{(\pm)}_a:=S_a{}_i{}^k\sigma^{\pm}{}_k{}^i \qquad \mbox{at all $a$,}
\end{equation}
where the quantities $S_a{}_i{}^k$ are defined in \p{S-Z-def}, $\sigma^{\pm}=\sigma^{1}\pm i\sigma^{2}$ and
$\sigma^{1,2}$ are the Pauli matrices.
So the conditions  \p{red-Z} lead to zero two generators in all $u(2)$ algebras \p{S-Z-def}, \p{S-S-Dir}.

After reduction with the conditions \p{red-Z} the obtained system involves only half of the initial semi-dynamical variables
\begin{equation}
\label{rest-Z}
z_a:= Z_a^{j=1}\,, \quad \bar z^a:=\bar Z^a_{j=1}\,, \qquad  \{ z_a, \bar z^b \}^{'}_{\scriptstyle{\mathrm{D}}} =-i\delta_a^b\,.
\end{equation}
Reduction of the Hamiltonian \p{Ham-fix} takes the form
\begin{equation}\label{Ham-fix-r}
\mathcal{H}  \ = \
\frac12\,\sum_{a}p_a p_a \ + \
\frac18\,\sum_{a\neq b} \frac{T_a{}^bT_b{}^a}{\sinh^2 \Big({\displaystyle\frac{q_a-q_b}{2}}\Big)}
\ - \, \frac{1}{8}\ {\rm Tr} \Big( \{\Phi^i, \bar\Phi_i\}\{ \Phi^k, \bar\Phi_k\} \Big)
\,,
\end{equation}
where
\begin{equation}\label{T-def-r}
T_a{}^b := z_a\bar z^b- \cosh\left(\frac{q_a-q_b}{2}\right)\{ \Phi^k, \bar\Phi_k \}_a{}^b\,.
\end{equation}
In this case, the $\mathcal{N}{=}\,4$ supersymmetry generators \p{Q} take the form
\begin{equation}\label{Q-r}
\begin{array}{rcl}
\mathcal{Q}^{\,i} &=& {\displaystyle  \sum\limits_{a} p_a \Phi^i{}_a{}^a \ -\ \frac{i}{2}\sum\limits_{a\neq b}
\frac{ T_a{}^b \Phi^i{}_b{}^a}{\sinh\Big({\displaystyle\frac{q_a-q_b}{2}}\Big)}
\ +\ \frac{i}{2}\sum\limits_{a, b}\, [\Phi^k, \bar\Phi_k]_a{}^b\Phi^i{}_b{}^a\,, } \\ [8pt]
\bar{\mathcal{Q}}_{\,i} &=& {\displaystyle  \sum\limits_{a} p_a \bar\Phi_i{}_a{}^a\ -\ \frac{i}{2}\sum\limits_{a\neq b}
\frac{ T_a{}^b \bar\Phi_i{}_b{}^a}{\sinh\Big({\displaystyle\frac{q_a-q_b}{2}}\Big)}
\ -\ \frac{i}{2}\sum\limits_{a, b}\, [\Phi^k, \bar\Phi_k]_a{}^b\bar\Phi_i{}_b{}^a\,, }
\end{array}
\end{equation}
while the first class constraints \p{F-constr-d} become
\begin{equation}\label{F-constr-d-r}
\mathcal{F}_a := T_a{}^a -c= z_a\bar z^a- \{ \Phi^k, \bar\Phi_k \}_a{}^a - c\approx 0\qquad\mbox{(no summation over $a$)}\,.
\end{equation}

Similarly to quantities \p{T-def} with the Dirac brackets \p{R-alg}, quantities \p{T-def-r} satisfy
\begin{equation}\label{T-alg}
\begin{array}{rcl}
\{T_a{}^b, T_c{}^d \}^{'}_{\scriptstyle{\mathrm{D}}}&=& -i\Big(\delta_a^d T_c{}^b-\delta_c^b T_a{}^d\Big) \\
&& -i\,
\sinh\Big({\displaystyle\frac{q_a-q_b}{2}}\Big)\sinh\Big({\displaystyle\frac{q_c-q_d}{2}}\Big)
\Big(\delta_a^d\{ \Phi^k, \bar\Phi_k \}_c{}^b-\delta_c^b\{ \Phi^k, \bar\Phi_k \}_a{}^d\Big)\,,
\end{array}
\end{equation}
As result, the charges \p{Q-r}, \p{Ham-fix-r} form the same ${\mathcal N}{=}\,4$ superalgebra \p{DB-QQ}-\p{DB-HQ},
up to the first class constraints \p{F-constr-d-r}.

However this reduced system contains $n$ first class constraints \p{F-constr-d-r} which,
together with the gauge fixing conditions, can eliminate all $n$ complex semi-dynamical variables $z_a$.
So similaly to the $\mathcal{N}{=}\,2$ case considered in \cite{Fed20}, we can make the gauge-fixing
\begin{equation}\label{fix-z}
\bar z^a= z_a \qquad\mbox{(for all $a$)}
\end{equation}
for the first class constraints \p{F-constr-d-r}.
Then, the components of the spinor $z_a$ become real and are expressed  through the remaining variables by the following expressions:
\begin{equation}\label{Z-sqrt}
z_a= \sqrt{c+\{ \Phi^k, \bar\Phi_k \}_a{}^a}   \qquad\mbox{(no summation over $a$)}\,.
\end{equation}
In this gauge the supercharges \p{Q1}, \p{bQ1} take the form
\begin{eqnarray}\nonumber
\mathcal{Q}^{\,i} &= &\sum\limits_{a} p_a \Phi^i{}_a{}^a-\frac{i}{2}\sum\limits_{a\neq b}
\frac{ \sqrt{c+\{ \Phi^k, \bar\Phi_k \}_a{}^a}\,\sqrt{c+\{ \Phi^j, \bar\Phi_j \}_b{}^b} \ \Phi^i{}_b{}^a}{\sinh\Big({\displaystyle\frac{q_a-q_b}{2}}\Big)} \\
&&  +\,\frac{i}{2}\sum\limits_{a\neq b}
\coth \Big({\displaystyle\frac{q_a-q_b}{2}}\Big)
\{ \Phi^k, \bar\Phi_k \}_a{}^b\,\Phi^i{}_b{}^a +\frac{i}{2}\sum\limits_{a, b} [\Phi^k{}, \bar\Phi_k]_a{}^b \Phi^i{}_b{}^a\,,
\label{Q2}\\
\nonumber
\bar{\mathcal{Q}}_{\,i} &= &\sum\limits_{a} p_a \bar\Phi_i{}_a{}^a-\frac{i}{2}\sum\limits_{a\neq b}
\frac{ \sqrt{c+\{ \Phi^k, \bar\Phi_k \}_a{}^a}\,\sqrt{c+\{ \Phi^j, \bar\Phi_j \}_b{}^b} \
\bar\Phi_i{}_b{}^a}{\sinh\Big({\displaystyle\frac{q_a-q_b}{2}}\Big)} \\
&&  +\,\frac{i}{2}\sum\limits_{a\neq b}
\coth\Big({\displaystyle\frac{q_a-q_b}{2}}\Big)
\{ \Phi^k, \bar\Phi_k \}_a{}^b\,\bar\Phi_i{}_b{}^a -\frac{i}{2}\sum\limits_{a, b} [\Phi^k{}, \bar\Phi_k]_a{}^b \bar\Phi_i{}_b{}^a\,.
\label{bQ2}
\end{eqnarray}
Moreover, in this gauge and in a pure bosonic limit,
the reduced Hamiltonian \p{Ham-fix-r} takes the form
\begin{equation}\label{Ham-fix-r-b}
\mathcal{H}_{bose}  \ = \
\frac12\,\sum_{a}p_a p_a \ + \
\frac18\,\sum_{a\neq b} \frac{c^2}{\sinh^2 \Big({\displaystyle\frac{q_a-q_b}{2}}\Big)}
\end{equation}
and is the Hamiltonian of the standard spinless  hyperbolic Calogero-Sutherland system.
Thus, the reduction \p{red-Z} of the considered system yields
gauge formulation of the  $\mathcal{N}{=}\,4$ spinless  hyperbolic Calogero-Sutherland system \cite{C,Su,OP,Poly-rev}.

Due to the presence of the square roots in the second terms in the supercharges \p{Q2}, \p{bQ2}
they contain higher degrees with respect to the Grassmannian variables.
To avoid this,  new variables
\begin{equation}\label{xi-def}
\xi^i{}_a{}^b= \Phi^i{}_a{}^b\sqrt{\frac{c+\{ \Phi^j, \bar\Phi_j \}_b{}^b}{c+\{ \Phi^k, \bar\Phi_k \}_a{}^a}}\,,  \qquad
\bar\xi_i{}_a{}^b= \bar\Phi_i{}_a{}^b\sqrt{\frac{c+\{ \Phi^j, \bar\Phi_j \}_b{}^b}{c+\{ \Phi^k, \bar\Phi_k \}_a{}^a}}
\end{equation}
were introduced in \cite{KLS-18b}.
In these quantities the supercharges \p{Q1}, \p{bQ1} take the form
\begin{eqnarray}\nonumber
{\mathcal{Q}}^{\,i} &= &\sum\limits_{a} p_a \xi^i{}_a{}^a -\frac{i}{2}\sum\limits_{a\neq b}
\frac{ \left(c+\{ \xi^k, \bar\xi_k \}_b{}^b\right) \xi^i{}_b{}^a}{\sinh\Big({\displaystyle\frac{q_a-q_b}{2}}\Big)} \\
&& +\,\frac{i}{2}\sum\limits_{a\neq b}
\coth \Big({\displaystyle\frac{q_a-q_b}{2}}\Big)
\{ \xi^k, \bar\xi_k \}_a{}^b\,\xi^i{}_b{}^a -\frac{i}{2}\,\beta\sum\limits_{a, b}[ \xi^k, \bar\xi_k ]_a{}^b\,\xi^i{}_b{}^a\,,
\label{Q3}\\
\nonumber
\bar{\mathcal{Q}}_{\,i} &= &\sum\limits_{a} p_a \bar\xi_i{}_a{}^a-\frac{i}{2}\sum\limits_{a\neq b}
\frac{ \left(c+\{ \xi^k, \bar\xi_k \}_b{}^b\right)
\bar\xi_i{}_b{}^a}{\sinh\Big({\displaystyle\frac{q_a-q_b}{2}}\Big)} \\
&& +\,\frac{i}{2}\sum\limits_{a\neq b}
\coth\Big({\displaystyle\frac{q_a-q_b}{2}}\Big)
\{ \xi^k, \bar\xi_k \}_a{}^b\,\bar\xi_i{}_b{}^a +\frac{i}{2}\,\beta\sum\limits_{a, b}  [ \xi^k, \bar\xi_k ]_a{}^b\,\bar\xi_i{}_b{}^a\,,
\label{bQ3}
\end{eqnarray}
where $\beta\,{=}\,{-}1$,
and coincide exactly with the ${\mathcal N}{=}\,4$ supersymmetry generators presented in \cite{KL-20}.\footnote{The author thanks Sergey Krivonos for the information that the value $\beta\,{=}\,{-}1$ is also valid in the hyperbolic case of the model presented in \cite{KL-20}.}
Point out that in contrast to the properties of the Grassmannian variables \p{Phi-def},
quantities \p{xi-def} do not form pairs with respect to complex conjugation,
that is some obstacle in quantization of the system in such representation.

\setcounter{equation}{0}
\section{Concluding remarks and outlook}

In this paper, the Hamiltonian description
of the $\mathcal{N}{=}\,4$ supersymmetric multi-particle hyperbolic Calogero-Sutherland system is presented,
which was obtained from the matrix superfield model by the gauging procedure \cite{FIL19}.
In contrast to the $\mathcal{N}{=}\,2$ case, the $\mathcal{N}{=}\,4$ supersymmetric generalization
of the gauged model
has the $\mathrm{U}(2)$ spin hyperbolic Calogero-Sutherland system as a bosonic core.

In the presented paper,
there are obtained explicit expressions of the $\mathcal{N}{=}\,4$ supersymmetry generators for different descriptions
of the system under consideration.
The supercharges \p{Q-matrix} and the Hamiltonian \p{Ham-matrix1} of the fully matrix system have a simple form,
but this system contains a large number of auxiliary degrees of freedom,
which can be eliminated by $n^2$ first class constraints \p{F-constr1}.
After the partial gauge fixing \p{x-fix}, eliminating off-diagonal even matrix variables,
we obtain the formulation in which the $\mathcal{N}{=}\,4$ supersymmetry generators \p{Q1}, \p{bQ1}
have the Calogero-like form and are closed on the Hamiltonian \p{Ham-fix} (or \p{Ham-fix1}) and $n$ first class constraints \p{F-constr-d}
generating the residual $[\mathrm{U}(1)]^n$ gauge symmetry.
Without off-diagonal odd variables in the classical supercharges \p{Q} (or \p{Q1}, \p{bQ1}), the nontrivial interaction terms disappear in them.

It is possible to impose the reduction conditions \p{red-Z}
that are $\mathcal{N}{=}\,4$ supersymmetry invariant and eliminate half of the spinning variables.
As result, we get the $\mathcal{N}{=}\,4$ supersymmetric system with $n$ first class constraints \p{F-constr-d-r},
which allows us gauging of the remaining spinning variables. Such a reduced system is in fact
the $\mathcal{N}{=}\,4$ generalization of the spinless hyperbolic Calogero-Sutherland system
equivalent to the model  presented in \cite{KL-20}.

In addition, the Lax representation \p{L-eq}, \p{Ps-eq}, \p{Z-eq} of the equations of motion
for the system under consideration is presented.
The set of conserved quantities \p{F-conser}, \p{Ik-def}, \p{Z-inv} is found.
Analysis of the classical integrability of the $\mathcal{N}{=}\,4$ system considered here will be the subject of the next paper.

Moreover, a further research will be devoted to quantum integrability
of the supersymmetric $\mathcal{N}{=}\,2$ and $\mathcal{N}{=}\,4$ systems constructed here.
Supersymmetry quantum generators are obtained using the Weyl ordering in quantum analogs of quantities such as
the $\mathcal{N}{=}\,2$ supersymmetric case.
However, in contrast to the $\mathcal{N}{=}\,2$ case \cite{Fed20},
due to the $\mathrm{SU}(2)$-doublet nature of odd variables in the $\mathcal{N}{=}\,4$ case,
the separation of the invariant sector with only diagonal odd variables does not work in the $\mathcal{N}{=}\,4$ quantum case.

\smallskip
\section*{Acknowledgements}
I  would  like  to  thank  Evgeny Ivanov  and  Sergey Krivonos  for useful discussions.
This work was supported by the Russian Science Foundation, grant no.\,16-12-10306.

\end{document}